\documentclass[paper]{agujournal2019}
\usepackage{url}
\usepackage{lineno}
\usepackage{color}
\usepackage{booktabs}
\usepackage{amsmath}
\usepackage{amssymb}
\usepackage{graphicx}
\usepackage[export]{adjustbox}
\draftfalse

\newcommand{\ssr}{   {Space Sci. Rev. }}

\newcommand{\jgr}{   {J. Geophys. Res.}}
\newcommand{\grl}{   {Geophys. Res. Lett.}}
\newcommand{\apj}{   {Astrophys. J.}}
\newcommand{\apjl}{   {Astrophys. J. Lett.}}
\newcommand{\prl}{   {Phys. Rev. Lett.}}

\newcommand{\nat}{   {Nature}}
\newcommand{\aap}{   {Astronomy \& Astrophysics}}
\newcommand{\araa}{   {Annual Review of Astronomy and Astrophysics}}

\journalname{JGR: Space Physics}

\begin{document}

%% ---------------------------------------------------------------%%

\title{Relativistic electron precipitation events driven by solar wind impact on the Earth's magnetosphere}

\authors{Alexandra Roosnovo\affil{1}, Anton V. Artemyev\affil{2}, Xiao-Jia Zhang \affil{3,2}, Vassilis Angelopoulos\affil{2}, Qianli Ma \affil{4,5}, Niklas Grimmich\affil{6}, Ferdinand Plaschke\affil{6}, David Fischer\affil{7}, Magnes Werner\affil{7}}
\affiliation{1}{ISR Space Science and Applications Group, Los Alamos National Laboratory, Los Alamos, New Mexico, USA}
\affiliation{2}{Department of Earth, Planetary, and Space Sciences, University of California, Los Angeles, Los Angeles, California, USA}
\affiliation{3}{Department of Physics, University of Texas at Dallas, Richardson, Texas, USA}
\affiliation{4}{Department of Atmospheric and Oceanic Sciences, University of California, Los Angeles, Los Angeles, California, USA}
\affiliation{5}{Center for Space Physics, Boston University, Boston, Massachusetts, USA}
\affiliation{6}{Institut für Geophysik und Extraterrestrische Physik, Technische Universität Braunschweig, Braunschweig, Germany}
\affiliation{7}{Space Research Institute, Austrian Academy of Science, Graz, Austria}

\correspondingauthor{Alexandra Roosnovo}{aroosnovo@lanl.gov}

\begin{keypoints}
\item A CIR and ICME, both including interplanetary shocks, are observed resulting in relativistic electron precipitation from low-altitude orbit 
\item Duskside precipitation after CIR impact is driven by intense EMIC waves, showing distinct energy-$L$ dispersion from magnetic field distortion 
\item Dawnside precipitation after ICME impact is driven by intense whistler-mode waves resonating with electrons at high latitudes
\end{keypoints}

%\item Two cases of solar wind transient impact, one foremost comprising a CIR and the other an interplanetary shock, are observed resulting in low-altitude relativistic electron precipitation
%\item Dawn-side precipitation event is driven by intense whistler-mode waves resonating with electrons at high latitudes
%\item Dusk-side precipitation event is driven by intense EMIC waves and shows distinct energy-$L$-shell dispersion, likely generated by equatorial magnetic field distortion

\begin{abstract}
Certain forms of solar wind transients contain significant enhancements of dynamic pressure and may effectively drive magnetosphere dynamics, including substorms and storms. An integral element of such driving is the generation of a wide range of electromagnetic waves within the inner magnetosphere, either by compressionally heated plasma or by substorm plasma sheet injections. Consequently, solar wind transient impacts are traditionally associated with energetic electron scattering and losses into the atmosphere by electromagnetic waves. In this study, we show the first direct measurements of two such transient-driven precipitation events as measured by the low-altitude Electron Losses and Fields Investigation (ELFIN) CubeSats. The first event demonstrates storm-time generated electromagnetic ion cyclotron waves efficiently precipitating relativistic electrons from $>300$ keV to $2$ MeV at the duskside. The second event demonstrates whistler-mode waves leading to scattering of electrons from $50$ keV to $700$ keV on the dawnside. These observations confirm the importance of solar wind transients in driving energetic electron losses and subsequent dynamics in the ionosphere. 
\end{abstract}

\section{Introduction}

The dynamics of the Earth's magnetosphere, especially those of the Earth’s inner magnetosphere, are largely controlled by solar wind impacts \cite{book:Kivelson&Russell95}. The most intense and sudden types of impact are those which include interplanetary (IP) shock waves, which result from the interaction of fast and slow solar wind streams and manifest as the upstream shock structures accompanying the larger geoeffective solar wind transient phenomena, such as interplanetary coronal mass ejections (ICMEs) and corotating interaction regions (CIRs) \cite{Gopalswamy03,Nitta21,Gosling96:cir,Heber99:cir,Richardson18:cir}. Such impacts have the ability to trigger rapid, large-scale redistribution of energetic particle fluxes in the radiation belts \cite<e.g.,>{Blake97, Lyons05, Tsurutani95:solarwind,Tsurutani11:jastp}. This redistribution involves significant adiabatic effects related to magnetic field reconfiguration, as well as kinetic effects related to plasma wave generation and energetic particle scattering.

The impact of the strongly intensified solar wind dynamic pressure that is characteristic to the large-scale solar wind transients (sometimes seen as distinct pulses of augmented pressure) compresses the Earth’s dayside magnetosphere and has an immediate influence on charged particle dynamics. This includes the formation of unstable (anisotropic) particle velocity distributions \cite<e.g.,>[and references therein]{Zhao22:drift&swpressure} as well as electron flux dropouts and enhancements \cite<e.g.,>{Ma21:dynamics_pressure_decrease,DaSilva23:chorus_cme}. The basic mechanism for the formation of unstable particle distributions consists of the adiabatic heating of ions and electrons via induction electric fields. Such heating is usually more effective for equatorial particles, resulting in the formation of perpendicularly anisotropic particle populations which are unstable to whistler-mode waves \cite<see>{Sagdeev&Shafranov61,Kennel66} and electromagnetic ion cyclotron (EMIC) waves \cite<see, e.g.,>{Liu22:emic&solarwind,Yan23:emic&IS,Zuxiang23:ws_emic,Thorne&Kennel71}.

Indeed, in-situ spacecraft measurements have detected many cases of whistler-mode chorus \cite<e.g.,>{Zhou15:chorus&shock,Zhou23:chorus_pressure} and EMIC wave \cite<e.g.,>{Usanova12} generation in response to solar wind dynamic pressure increases, e.g., during an interplanetary shock wave’s arrival to the Earth’s magnetosphere. A detailed multi-case study by \citeA{Yue17:whistler&solarwind} demonstrated that IP shock impact can significantly increase the intensity of whistler-mode chorus waves in the outer radiation belt, outside of the plasmapause. Although this type of wave intensity enhancement is typical for any positive pulses (i.e., increases) of the solar wind dynamic pressure, IP shocks often provide the strongest effect \cite{Jin22:chorus&shock}. Interestingly, wave intensity increases not only around the equatorial plane, where the chorus generation region is located \cite<see reviews by>[and references therein]{Tao20,Omura21:review}, but in low-altitude regions as well \cite{Bezdekova21:waves&shocks}. This suggests that the more intense whistler-mode waves driven by IP shock impact are not damped by suprathermal electron fluxes \cite{Bortnik07:landau,Chen13} and can propagate to high latitudes, thus significantly increasing their global efficiency in scattering relativistic electrons \cite<see discussion in>{Chen21:frontiers,Chen22:microbursts,Artemyev21:jgr:ducts}.

Magnetospheric impact by strong solar wind transient IP shocks plays a similarly significant role in the intensification of EMIC waves \cite{Yan23:emic&IS}. \citeA{Blum21:emic&shock} described a CME event that led to a series of compressions of the dayside magnetosphere by pulses of solar wind dynamic pressure; each of such compressions resulted in proton adiabatic heating and near-equatorial EMIC wave generation. The effects of EMIC wave generation due to IP shock impact on the Earth’s magnetosphere can be even more evident, such as in the stark ion flux enhancements observed by \citeA{Li22:emic&shock} and \citeA{Zuxiang23:ws_emic}. Moreover, EMIC wave intensity enhancements in response to solar wind pulses may also be observed simultaneously on the day and night sides of the Earth, as when coinciding with plasma sheet ion injections driven by substorm activities \cite{Xue22:emic&swpressure,Yan23:emic&IS}. For EMIC wave generation in particular, the solar wind impact may consist of two independent processes: (1) direct proton heating by magnetic field compression within the inner magnetosphere and (2) injection of hot, anisotropic protons into the inner magnetosphere by flow bursts and dipolarizing flux bundles arising from localized reconnection in the magnetotail  \cite<see discussion and comparison of these two processes in, e.g.,>{Chen20:emic_statistics,Upadhyay22}.

Although whistler-mode and EMIC wave generation caused by interplanetary shock waves and solar wind dynamic pressure pulses has been previously reported,  as it is commonly observed by near-equatorial spacecraft, details on the influence of these waves on radiation belt dynamics have yet to be fully investigated. One expected effect of importance is the scattering and resultant precipitation of energetic electrons by intense whistler-mode and EMIC waves. However, such electron precipitation can only be observed by low-altitude spacecraft (i.e., taking advantage of finite, $\sim 20$ deg, pitch-angle resolution measurements of electron distributions within a large, i.e., many tens of degrees, loss-cone) or ground-based measurements of X-ray emission \cite<see, e.g., example in>{Breneman20:losses&swpressure}. Direct measurements of precipitating electron fluxes in response to solar wind dynamic pressure enhancements can therefore be highly useful for understanding the importance of transients, including IP shocks, in magnetosphere-ionosphere coupling and radiation belt depletion.

Here, we describe two events in which large-scale solar wind structures impact the Earth’s magnetosphere and drive relativistic electron losses. Both precipitation events were captured by the low-altitude measurements of the Electron Losses and Fields Investigation (ELFIN) CubeSats \cite{Angelopoulos20:elfin}. The first event consists of a magnetospheric impact by a CIR (with an embedded IP shock and prominent solar wind discontinuities) that drives a magnetospheric storm and strong relativistic electron precipitation by EMIC waves on the duskside; the second event consists of a separate ICME impact, adjoined by a prominent IP shock, that drives strong energetic electron precipitation, extending to relativistic energies, by whistler-mode waves on the dawnside. We describe the satellite observations of the solar wind, inner magnetosphere, and low-altitude space region in Section \ref{sec:observations}. In Sections \ref{sec:1} and \ref{sec:2}, we examine the first and second events, respectively. In Section \ref{sec:discussion}, we discuss our results and the likely characteristics of the specific waves responsible for the two electron precipitation events. Finally, we summarize our results and present conclusions in Section \ref{sec:conclusions}.

\section{Observations}\label{sec:observations}
We examine two specific events of electron precipitation, observed from the low-altitude vantage point of ELFIN, driven by interplanetary shock interaction with the terrestrial magnetosphere: the first event (S\#1) occurred on 6 March 2021, and the second event (S\#2) occurred on 12 May 2021. We use ELFIN observations of precipitating (inside the local bounce loss-cone) and locally trapped (outside the local bounce loss-cone) fluxes within the energy range of 50-6000 keV (16 energy channels) at 3 s time resolution (ELFIN spin rate) \cite{Angelopoulos20:elfin}. We also use the precipitating-to-trapped flux ratio as an effective measure of the intensity of electron precipitation \cite<see examples in>{Mourenas21:jgr:ELFIN,Tsai22,Zhang22:natcom}. 

\begin{figure*}
\centering
\includegraphics[width=.95\textwidth]{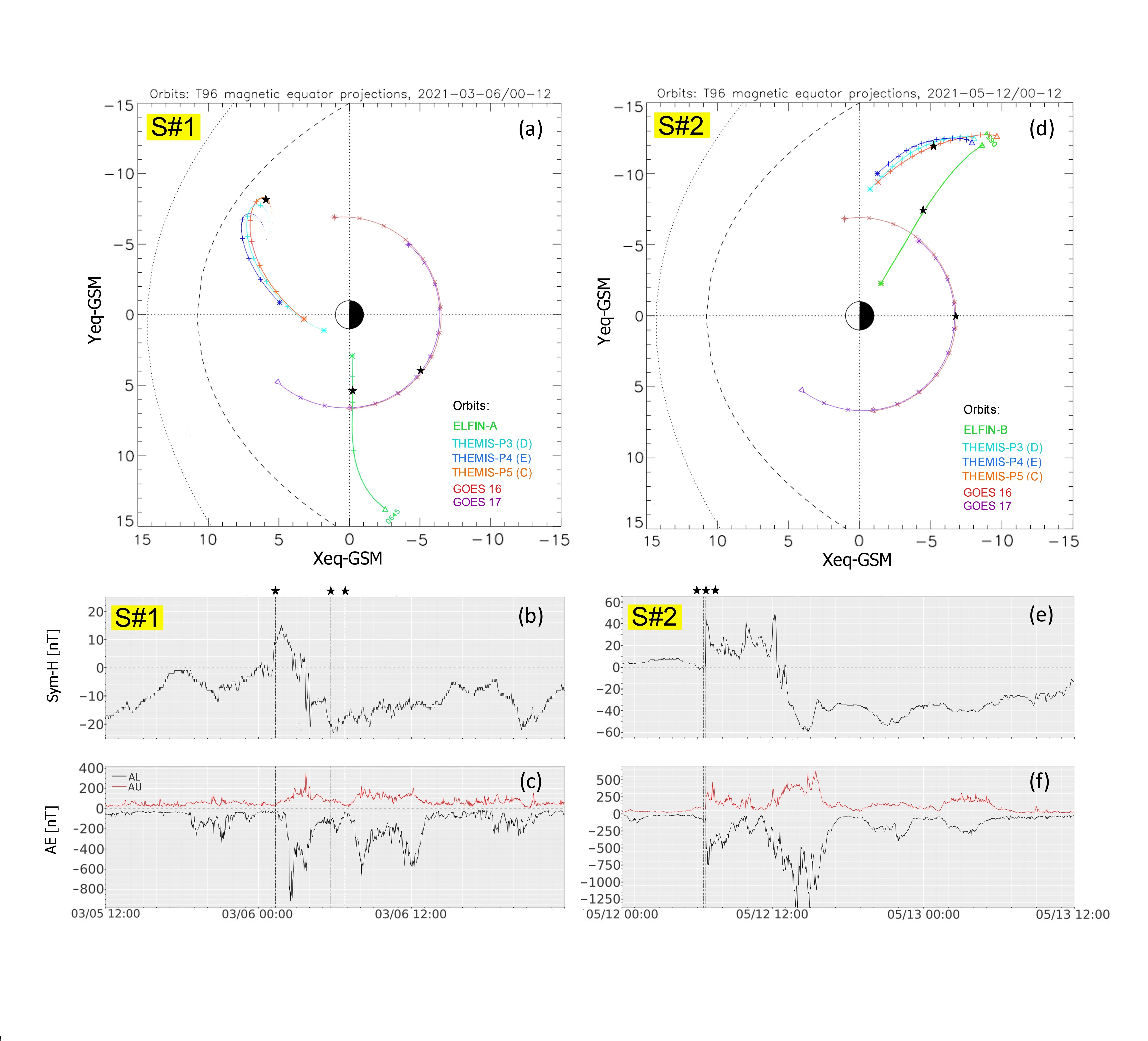}
\caption{Equatorially-projected positions of THEMIS, GOES, and ELFIN spacecraft (a, d), relative to the modelled nominal magnetopause (dashed curve) and bow shock (dotted curve). ARTEMIS (THEMIS B and C) is located in the solar wind, out of frame. For each orbit, the start time is marked with a triangle while the end time is marked with an asterisk; the tick marks between represent hour intervals for THEMIS and GOES satellites and minute intervals for ELFIN. The bottom panels show Sym-H (b, e) and AE (c, f) indexes for the two-day interval encompassing each event. The orbits of observation and geomagnetic indices for the first event (S\#1) are shown in the left panels, while those of the second event (S\#2) are shown on the right. The stars mark the approximate locations and times for the different shock observations made by ARTEMIS, THEMIS, and ELFIN.
\label{fig1} }
\end{figure*}

To monitor the solar wind and magnetospheric conditions for perturbations indicative of transient and accompanying shock arrival, we utilize observations from the Time History of Events and Macroscale  Interactions during Substorms (THEMIS) mission and The Geostationary Operational Environmental Satellite (GOES). Measurements of the upstream solar wind, where an approaching IP shock is first observable as a sharp gradient of solar wind velocity and magnetic field magnitude, are taken from the Acceleration, Reconnection, Turbulence, and Electrodynamics of the Moon’s Interaction with the Sun (ARTEMIS) subset of THEMIS spacecraft (specifically, ARTEMIS P2 also know as THEMIS C, with the latter designation utilized hereafter). The ARTEMIS satellites orbit the moon and measure the solar wind magnetic field \cite{Auster08:THEMIS} and plasma \cite{McFadden08:THEMIS,Artemyev18:jgr:report}. The three other THEMIS spacecraft (A, D, and E) orbit the Earth with an apogee of $\sim 12R_E$ \cite{Angelopoulos08:ssr}. We use THEMIS A magnetic field and plasma measurements ( 3-4 s spin resolution) to monitor the near-Earth dayside magnetosheath and foreshock response to the arriving interplanetary shock. Additionally, to identify plasma injections we check energetic ion and electron measurements made by the GOES-16 and GOES-17 space weather suite of instruments \cite{Dichter15:goesr,Boudouridis20:goesr}. Figure \ref{fig1} shows the orbits of THEMIS, GOES, and ELFIN spacecraft relative to the nominal, modelled magnetopause and bow shock \cite{Shue97:magnetopause,Wu00:bowshock,King&Papitashvili05}, as well as the geomagnetic activity, as represented by Sym-H and AE indices, around the time of each event. 

\subsection{First event: EMIC wave-driven precipitation}\label{sec:1}
\begin{figure*}
\centering

\includegraphics[width=0.95\textwidth]{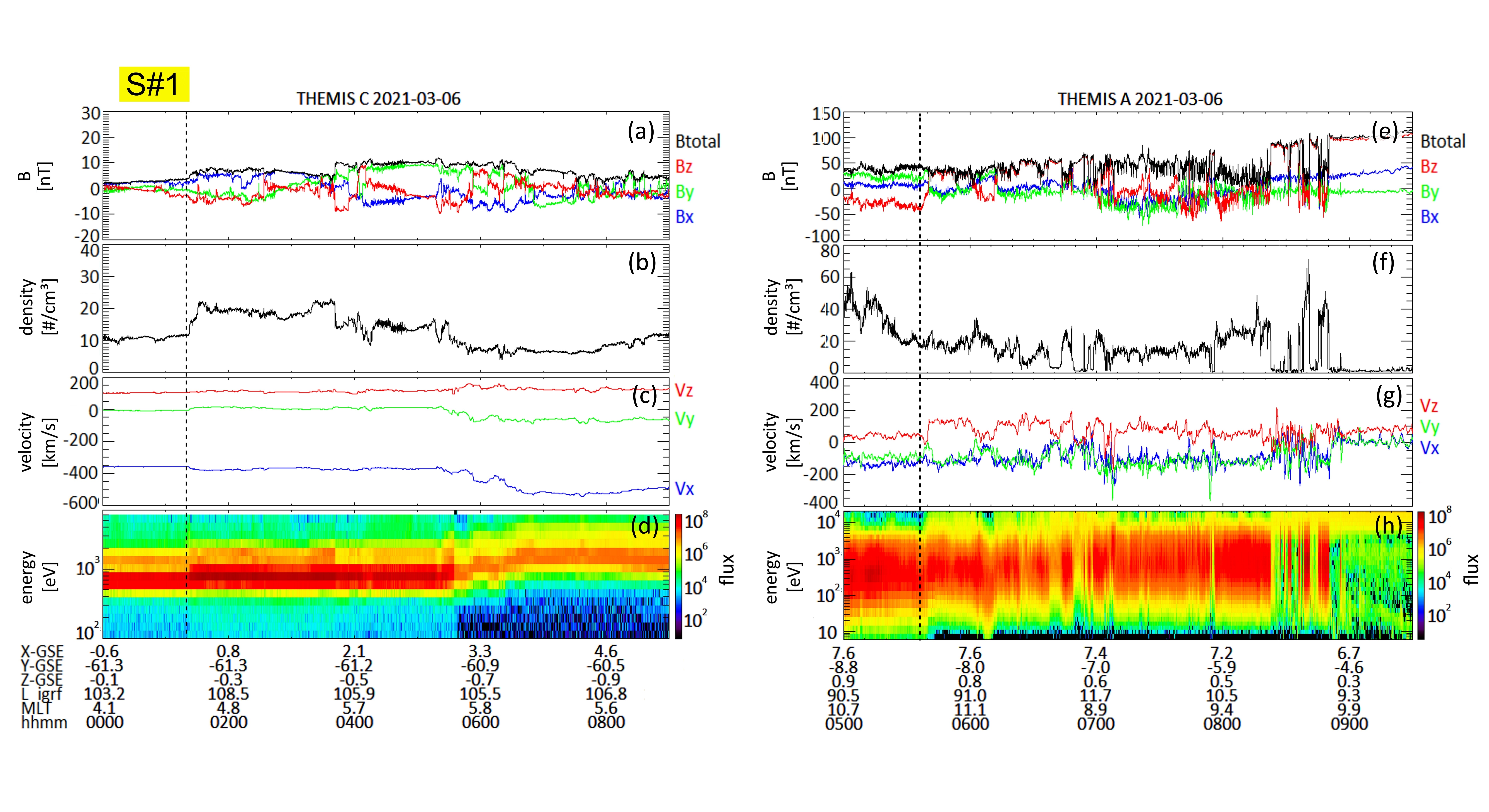}
\caption{Overview of ARTEMIS (THEMIS C) and THEMIS A observations for event 1 on 6 March 2021: THEMIS C magnetic field (a), plasma density (b), plasma flow speed (c), and ion energy spectrum (d) and THEMIS A magnetic field (e), plasma density (f), plasma flow speed (g), and ion energy spectrum (h), with the colorbar showing flux in [cm$^{-2}$s$^{-1}$sr$^{-1}$eV$^{-1}$]. At the bottom of each set of panels (a-d, e-h) are location and time information, including the X, Y, and Z positions in the GSE coordinate system, L and MLT values, and the hour (hh) and minute (mm) for the day of the event. The beginning of the primary disturbances caused by the shock are indicated by the dashed lines across panels (a-d) and (e-h), as observed by THEMIS C and THEMIS A, respectively.
\label{fig2} }
\end{figure*}

%(2.2)
Our first event occurred on 6 March 2021. Figure \ref{fig2} shows an overview of ARTEMIS (THEMIS C) and THEMIS A observations. THEMIS C observes the large-scale solar wind perturbations of a corotating interaction region (CIR) \cite<see>{Gosling96:cir,Heber99:cir,Richardson18:cir}, starting at $\sim$ 01:00 UT with a slight jump of solar wind speed (panel (b)). Simultaneous variations of magnetic field magnitude (panel (a)) and plasma density (panel (b)) show the series of rotational discontinuities (rotation of ${\bf B}$ components with $|{\bf B}|\approx const$) associated with an interplanetary shock wave ($|{\bf B}|$ variation) embedded in a CIR \cite<see detailed discussion in>{Gosling96:cir}. Distinct from the initial fine structure of solar wind perturbations, the large scale magnetic field and solar wind discontinuities, seen prominently in the ion spectra variation around 05:40-06:00 UT in panel (c), are expected to compress the Earth's magnetosphere and drive a geomagnetic storm \cite<see>{Gonzalez99:storms,Alves06}. Indeed, Sym-H and AE indexes in Figure \ref{fig1} show moderately depressed Sym-H, indicative of storm-like activity, along with so-called high-intensity, long-duration, continuous AE activity, \cite<as described in>{Tsurutani04, Tsurutani06}. Activity starts with magnetosphere compression from 01:00-02:00 UT (positive Sym-H) and continues to a moderately negative Sym-H of around $-20$ nT with recurrent substorms (AL minimum reaching $\sim -800$ nT). The substorm around 03:00 UT is associated with a strong ion injection observed at GOES-16 in the pre-midnight sector (not shown). Such events are usually characterized by an increased level of relativistic electrons in the inner magnetosphere \cite{Hajra14,Hajra15}, but have not been studied in the context of relativistic electron precipitation. 

During THEMIS C observations of the CIR, THEMIS A, located inside the magnetosheath (see Figure \ref{fig1}), detects multiple strong magnetic field perturbations accompanied by density variations and hot magnetospheric plasma bursts. Such variations of density and cold/hot plasma flux are indicative of multiple magnetopause crossings due to magnetopause surface waves \cite<e.g.,>{Agapitov09,Archer19} or Kelvin-Helmholtz waves \cite<e.g.,>{Hasegawa04}. $B_z$ changes sign multiple times, i.e., the magnetosheath is filled by negative polarity $B_z$ variations that potentially drive magnetopause reconnection \cite{Paschmann79,Paschmann13,Phan14,Burch16:science}; multiple plasma jets ($v_z$ excursions from the ambient sheath flow) are also seen. Thus, THEMIS A confirms the strong driving of the Earth's magnetosphere by the CIR after its arrival at $\sim$01:00 UT.

\begin{figure*}
\centering
\includegraphics[width=0.95\textwidth]{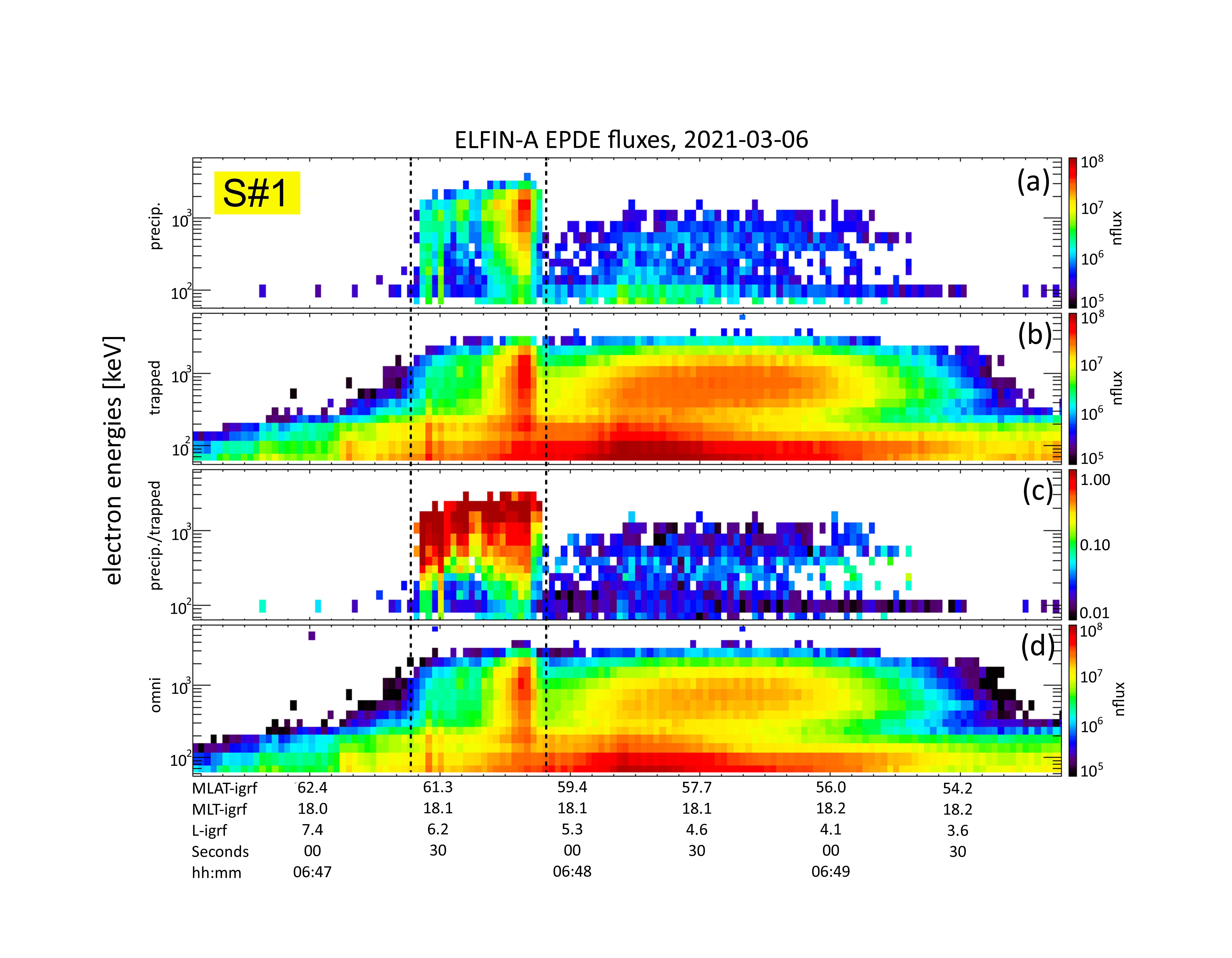}
\caption{Overview of ELFIN observations for event 1 on 6 March 2021: precipitating electron fluxes (a), trapped fluxes (b), precipitating-to-trapped flux ratios (c), and omnidirectional fluxes (d). In panels (a), (b), and (d) the colorbar shows flux in [cm$^{-2}$s$^{-1}$sr$^{-1}$MeV$^{-1}$]. The dashed lines demarcate the time interval in which electron precipitation is primarily observed, as indicated in the enhancement of the precipitating-to-trapped ratio (i.e., ratio approaches unity).
\label{fig3} }
\end{figure*}

At $\sim$06:47 UT, near the time of the observed Sym-H minimum (1 hour after THEMIS C detects the ending edge of the CIR with a large increase of the solar wind speed), and still well within the prolonged, albeit weak, storm main phase, ELFIN A crosses the dusk flank (MLT$\sim18$) and observes strong precipitation of relativistic electrons. Figure \ref{fig3} shows an overview of flux observations capturing this precipitation. The precipitation burst covers a wide range of magnetic latitudes, $MLAT\in[61.4,59.4^\circ]$ (corresponding to a wide $L$-shell range in the equatorial region of electron scattering, $\Delta L\sim 1$),  between the plasma sheet region (before 06:47:30 UT; region with only $<300$ keV electron fluxes; see detailed analysis of such ELFIN observations in, e.g., \citeA{Artemyev22:jgr:ELFIN&THEMIS}) and plasmasphere (after 06:48:30 UT; region with characteristic depletion of $\sim 100-200$ keV fluxes due to scattering by plasmaspheric hiss waves; see detailed analysis of such ELFIN observations in, e.g., \citeA{Mourenas21:jgr:ELFIN}). The precipitating-to-trapped electron flux ratio maximizes at $>300$ keV and stays $\sim 1$ for energies up to $\sim 2-3$ MeV. Relativistic electron precipitation distinctly lacking in energies $<300$ keV is the characteristic feature of electron resonant scattering by EMIC waves \cite<see detailed description of such events in, e.g.,>{An22:prl,Grach22:elfin,Capannolo23:elfin,Angelopoulos23:ssr}, as the minimum resonance energy for such scattering is typically $\gtrsim 0.5$ MeV \cite{Summers&Thorne03,Kersten14,Ni15}. Meanwhile, GOES-17, located in the pre-midnight sector, observes a strong $B_z$ depletion of $\sim -50$ nT (not shown) that is associated with a ring current injected ion population \cite{Daglis99}, the principal source of EMIC waves \cite<e.g.,>{Chen10:emic,Chen11:emic}. The duskside location of the precipitation event further supports characterization as EMIC-driven scattering, as this is the primary region of EMIC wave generation, with aforementioned hot plasma sheet (ring current) ions drifting duskward after being injected at the nightside \cite{Thorne&Kennel71,Jun19:emic,Jun21:emic}. 

\begin{figure*}
\centering
\includegraphics[width=0.95\textwidth]{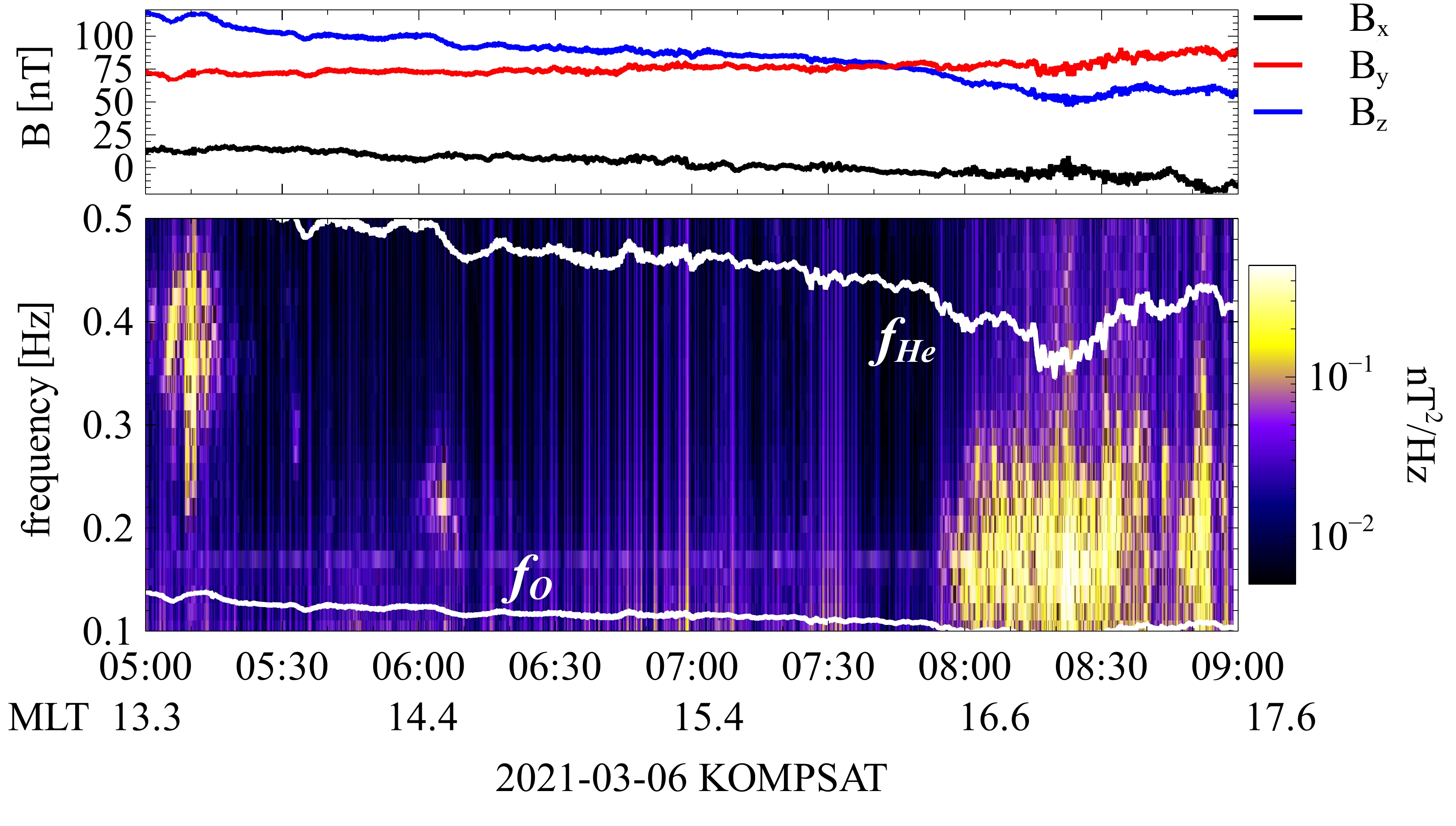}
\caption{An overview of KOMPSAT magnetic field measurements at the dusk flank around the time of event 1: magnetic field components (top panel) and magnetic field spectrum for EMIC wave frequency range (bottom panel). The two white lines depict helium (He) and oxygen (O) gyrofrequencies. 
\label{fig:kompsat} }
\end{figure*}

Although there was no direct magnetic conjunction of ELFIN with near-equatorial spacecraft during the first event, the geostationary GEO-KOMPSAT-2A \cite{Seon20:KOMPSAT} satellite was traveling along the dusk flank around the time of ELFIN electron precipitation observations and observed several intense bursts of helium band EMIC waves. Figure \ref{fig:kompsat} shows KOMPSAT fluxgate magnetometer measurements (1 s resolution; \citeA{Magnes20:KOMPSAT_fgm,Constantinescu20:KOMPSAT}) during the interval of 05:00-09:00 UT. There are clear EMIC wave bursts (bottom panel) around 05:10, 06:10, and 08:00-09:00 UT, covering an MLT range that extends from 13 up to 18 hours. These time intervals and MLT locations do not exactly overlap with ELFIN measurements at $\sim $06:50 UT, MLT$\sim 18$, but instead provide a good context for ELFIN measurements. KOMPSAT shows that a large part of the dusk flank is filled by EMIC wave source regions at the times surrounding our observations; these regions can survive for a long time and be quite extended in MLT \cite<see>{Engebretson15,Blum20}. Thus, the observation of multiple EMIC wave source regions in close spatial and temporal proximity to ELFIN observations of relativistic electron precipitation follows our interpretation of EMIC wave scattering.

Interestingly, EMIC wave-driven precipitation is quite long lasting (multiple ELFIN spins covering almost $\Delta L\sim 1$ and reaching the upper limit of the range of sizes expected for an equatorial EMIC wave source region \cite{Blum16,Blum17}). Our first event additionally includes two types of $dE/dL$ (or $dE/dMLAT$) gradients: one around 06:47:30 UT, with the minimum precipitating electron energy increasing as $L$-shell decreases ($dE/dL<0$), and a second around 06:47:45 UT, when the minimum precipitating electron energy decreases as $L$-shell decreases ($dE/dL>0$). The $dE/dL$ gradient is likely provided by the dependence of the minimum resonance energy on the equatorial ratio of the plasma frequency and the gyrofrequency, $E\propto  f_{ce}/f_{pe}$ \cite{Summers&Thorne03}. In the unperturbed dipole magnetic field $f_{ce}\propto  L^{-3}$ and $f_{pe}\propto  L^{-2}$ \cite<see the empirical model in>{Sheeley01}, which will give $E\propto L^{-1}$ with $dE/dL<0$ (observed at larger $L$, around 06:47:30 UT). Substorm injections, however, transport hot ion populations  \cite{Birn15,Gkioulidou14,Gkioulidou16,Ukhorskiy17:DF,Ukhorskiy18:DF} that  may form localized regions of magnetic field depletion \cite<so-called magnetic dips,>{Xia19:magnetic_hole,Zhu21:magnetic_holes} filled by EMIC waves \cite<see>{He17:emic,Yin22:magnetic_holes&emic,Yu23:magnetic_holes&emic,Zhao23:emic_dips}. This magnetic field depletion will result in a weaker radial gradient of $f_{ce}$, that is $f_{ce}\propto L^{-3+q}$ with $q>0$ \cite{Xia19:magnetic_hole,Zhu21:magnetic_holes}, and this effect may make  $f_{ce}/f_{pe}\propto L^{-1+q}$ increase with decreasing $L$ (i.e., $dE/dL>0$, observed around 06:47:45 UT). Therefore, the inverse gradient of the precipitating energies ($dE/dL>0$, seen in  Fig. \ref{fig3}) corroborates that here a strong ion injection is penetrating deep into the plasmapause and driving significant losses of relativistic electrons. Indeed, KOMPSAT magnetic field measurements around MLT $\sim 13.3$ and $\sim 17$ show a significant magnetic field depletion within the source region of EMIC waves (Figure \ref{fig:kompsat}, top panel; for MLT $\sim 17$, depletion is seen also in the gyrofrequency profile of the bottom panel below). 

\subsection{Second event: whistler-mode wave-driven precipitation}\label{sec:2}
\begin{figure*}
\centering
\includegraphics[width=0.95\textwidth]{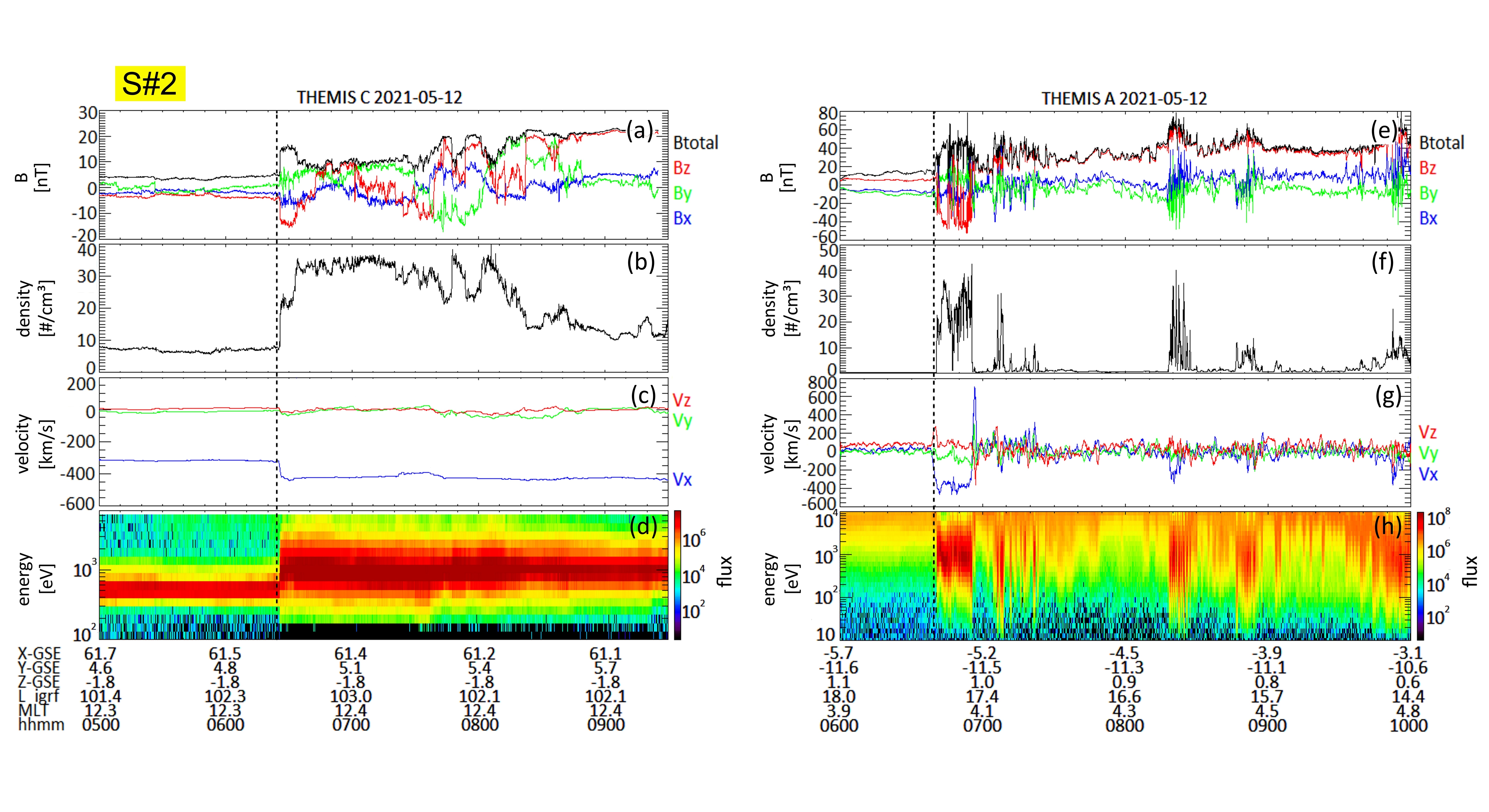}
\caption{Overview of ARTEMIS (THEMIS C) and THEMIS A observations for event 2 on 12 May 2021:  THEMIS C magnetic field (a), plasma density (b), plasma flow speed (c), and ion energy spectrum (d) and THEMIS A magnetic field (e), plasma density (f), plasma flow speed (g), and ion energy spectrum (h), with the colorbar showing flux in [cm$^{-2}$s$^{-1}$sr$^{-1}$eV$^{-1}$]. At the bottom of each set of panels (a-d, e-h) are location and time information, including the X, Y, and Z positions in the GSE coordinate system, L and MLT values, and the hour (hh) and minute (mm) for the day of the event. The beginning of the primary disturbances caused by the shock are indicated by the dashed lines across panels (a-d) and (e-h), as observed by THEMIS C and THEMIS A, respectively. 
\label{fig4} }
\end{figure*}

The second event occurred on 12 May 2021. THEMIS C, in the solar wind ahead of the Earth's bow shock, observes the IP shock of an impinging ICME \cite{Gopalswamy03,Nitta21} around 06:25 UT. Figure \ref{fig4}(a-d) shows the strong gradient of the magnetic field magnitude, a plasma density jump from 8 cm$^{-3}$ to $\sim$40 cm$^{-3}$, and an intensification of solar wind flow from $\sim-300$ km/s to $\sim -450$ km/s; the ion spectrum also shows conspicuous flow and thermal energy increases across the shock. Compared with the first event (S\#1, Figure \ref{fig2}), the IP shock of the second event is much more distinct in form, with sharper gradients between upstream and downstream regions.

Prior to the IP shock's arrival to the Earth's magnetosphere, THEMIS A was inside the magnetosphere and observed hot stagnant ions (i.e., ion energy is above $1$ keV and ion flow is around zero, see Figure \ref{fig4}(e-h)). The shock impact compresses the magnetosphere and moves the magnetopause toward the Earth, such that THEMIS A momentarily appears to be located within the magnetosheath, with high density plasma flow observed onward from 06:40 UT. The spacecraft returns to the magnetosphere around 07:00 UT, the magnetopause evidently moving out closer to its pre-shock configuration. However, THEMIS A undergoes multiple apparent magnetopause crossings over the subsequent $\sim3$ hours; such crossings are seen as plasma density increases along with alternating recurrences of hot rarefied and cold dense ion populations in the flux spectrum shown in Figure \ref{fig4}(h). These successive magnetopause crossings are likely caused by magnetopause oscillation, driven by both IP shock impact and the arrival of subsequent trailing solar transients that compose the extent of CME (observed by THEMIS C behind the initial IP shock) \cite<see, e.g.,>{Agapitov09,Archer19}. During the entire interval of 06:00-10:00 UT, THEMIS A was at $L\sim 14$; thus observations of magnetopause crossings after 07:00 UT highlight the large amplitude character of the magnetopause oscillations.

\begin{figure*}
\centering
\includegraphics[width=0.95\textwidth]{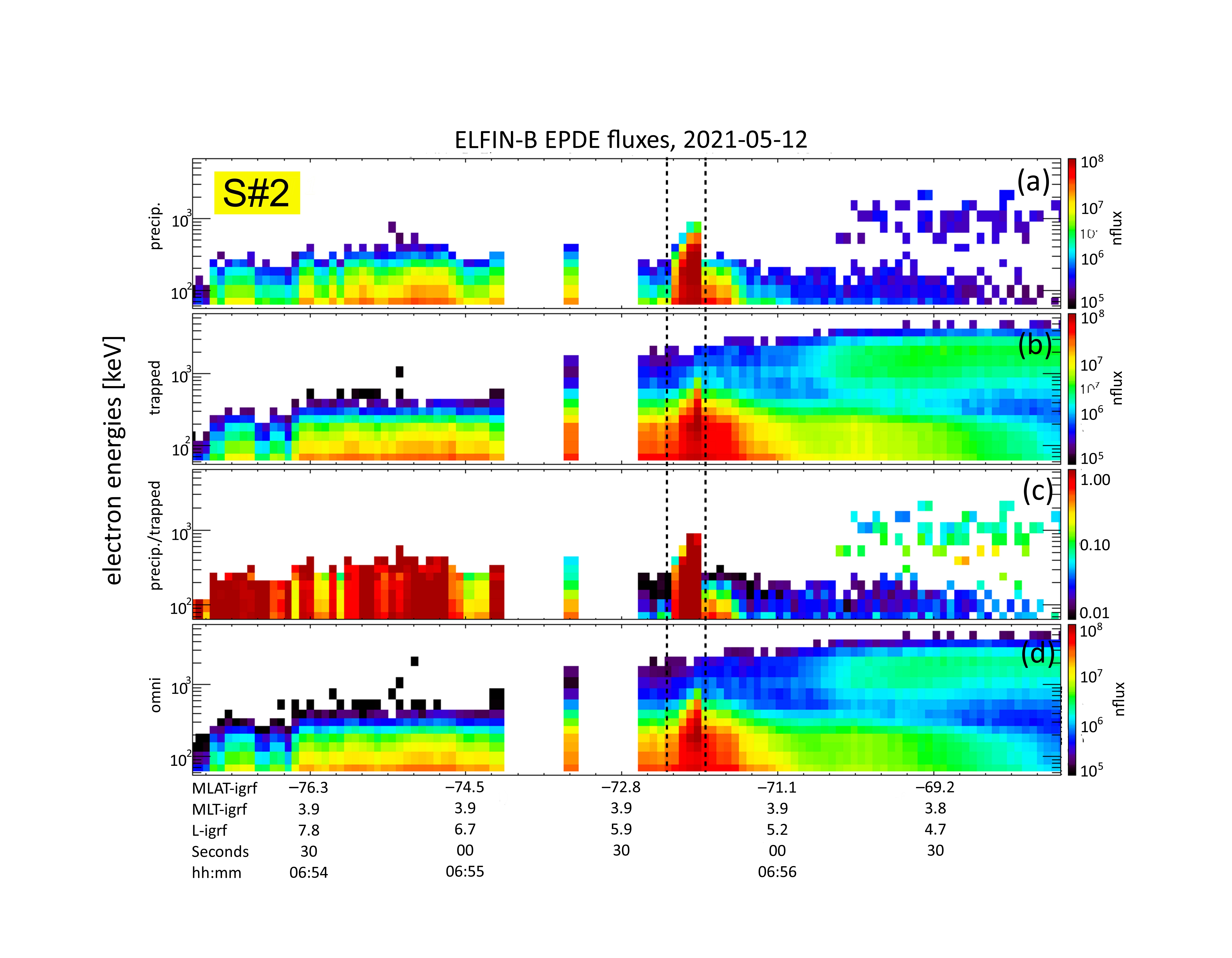}
\caption{Overview of ELFIN observations for event 2 on 12 May 2021: precipitating electron fluxes (a), trapped fluxes (b), precipitating-to-trapped flux ratios (c), and omnidirectional fluxes (d). In panels (a), (b), and (d) the colorbar shows flux in [cm$^{-2}$s$^{-1}$sr$^{-1}$MeV$^{-1}$]. The dashed lines demarcate the time interval in which electron precipitation is primarily observed, as indicated in the enhancement of the precipitating-to-trapped ratio (i.e., ratio approaches unity).
\label{fig5}}
\end{figure*}

Figure \ref{fig1} shows that the IP shock compresses the magnetosphere (evidenced by the long interval of increased, positive Sym-H for S\#2) and drives a substorm with AE$\sim -700$ nT (both GOES-16 and 17 observe strong plasma sheet injections at 06:40 UT on the nightside; not shown). Additionally, after 11:00 UT there are prolonged storm activities with Sym-H around $-60$ nT \cite<expected for CME impact; see>[and references therein]{Tsurutani03:cme,Koehn22}, similar to what we observe for the first event, albeit more intense here. Focusing on the initial compressing shock impact at 07:00 UT, we see ELFIN B crossing the dawn-flank magnetosphere when it observes a very intense burst of electron precipitation. Figure \ref{fig5} shows ELFIN detecting the relativistic electron precipitation burst around $L\sim 5.7$, with an upper energy of $\sim 800$ keV and the precipitating-to-trapped flux ratio reaching 1 for approximately the entire energy range. This burst is localized between the plasma sheet (distinguished by the absence of trapped fluxes > 300 keV and the presence of isotropic fluxes $<300$ keV for electrons observed before 06:55 UT) and the plasmapause (recognized by the disappearance of $\sim 300$ keV fluxes after 06:56 UT; see discussions of this feature in ELFIN observations by \citeA{Mourenas21:jgr:ELFIN,Angelopoulos23:ssr}). Therefore, this precipitation burst occurred in the outer radiation belt, where electrons are scattered by intense whistler-mode waves in the dawn region \cite<strong precipitation of $\sim50$ keV electrons is the defining characteristic of whistler-mode wave scattering, see, e.g.,>{Tsai22,Chen22:microbursts}. This precipitation burst covers four ELFIN spins, i.e., it lasts much longer than microburst precipitation duration \cite<e.g.,>{OBrien04,Shumko21}. Considering these features in totality, we interpret this instance of relativistic electron precipitation, localized within the outer radiation belt, as an equatorial intensification of whistler-mode waves due to IP shock-induced magnetospheric compression \cite<see, e.g.,>{Yue17:whistler&solarwind,Jin22:chorus&shock}.

\section{Discussion}\label{sec:discussion}
We have presented two different events with EMIC and whistler-mode wave-driven electron precipitation bursts (first event, S\#1, and second event, S\#2, respectively). Both events are characterized by strong solar wind drivers that either provide ion injections, followed by EMIC wave generation, or electron compressional heating, followed by whistler-mode wave generation. We now estimate the physical characteristics of EMIC and whistler-mode waves that would be required to obtain the types of electron precipitation spectra we observed for each event. We analyze two main aspects of wave generation: the resonance conditions and the cold plasma dispersion relation. Firstly, we look at these aspects for whistler-mode waves. The most intense whistler-mode waves propagate along magnetic field lines \cite{Li11,Agapitov13:jgr}, and in cold dense plasma their dispersion relation takes the following form \cite{bookStix62}:

\begin{equation}
\omega  = \Omega _{ce} \left( \lambda  \right) \cdot \left( {1 + \frac{{\Omega _{pe}^2 \left( \lambda  \right)}}{{k^2 \left( \lambda  \right)c^2 }}} \right) \label{eq01}    
\end{equation}
or
\[
k\left( \lambda  \right) = \frac{{\Omega _{pe} \left( \lambda  \right)}}{c} \cdot \left( {\frac{{\Omega _{ce} \left( \lambda  \right)}}{\omega } - 1} \right)^{ - 1/2} 
\]

where the dispersion relation sets the wave number $k(\lambda)$ for a fixed wave frequency, $\omega$. The electron gyrofrequency $\Omega_{ce}=\Omega_{ce,eq}\sqrt{1+3\sin^2\lambda}/\cos^6\lambda$ is given by the dipole magnetic field model ($\Omega_{ce,eq}$ is the equatorial gyrofrequency; $\lambda$ is the magnetic latitude), and the plasma frequency $\Omega_{pe}=\Omega_{pe,eq}\cos^{-5/2}\lambda$ is given by the \citeA{Denton06} model ($\Omega_{pe,eq}$ is the equatorial plasma frequency; the ratio $\Omega_{pe,eq}/\Omega_{ce,eq}$ is taken from the model presented in \citeA{Sheeley01}). The resonance condition for field-aligned whistler-mode waves is:
\begin{equation}
\gamma\omega  - k\left( \lambda  \right)p_\parallel\left( \lambda  \right)/m_e  = \Omega _{ce} \left( \lambda  \right) \label{eq02}    
\end{equation}
where electron parallel momentum $p_\parallel$ can be written as a function of electron energy $m_ec^2(\gamma-1)$ and equatorial pitch-angle $\alpha_{eq}$:
\begin{equation}
p_\parallel =  - m_e c\sqrt {\gamma ^2  - 1} \sqrt {1 - \sin ^2 \alpha _{eq} \frac{{\Omega _{ce} \left( \lambda  \right)}}{{\Omega _{ce,eq} }}} \label{eq03}    
\end{equation}
We are interested in electron precipitation, and thus the equatorial pitch-angle should be defined by the loss-cone size, $\alpha_{LC}\approx L^{-3/2}\cdot\left(4-3/L\right)^{-1/4}$, where $L$ (L-shell) is defined by the radial distance (in Earth radii) of the equatorial crossing of Earth's magnetic field lines \cite{bookSchulz&anzerotti74}. Combining the resonance condition (\ref{eq02}), dispersion relation (\ref{eq01}), and equation for $\alpha_{eq}=\alpha_{LC}$, we obtain the precipitating electron energy as a function of magnetic latitude for a given  $\Omega_{pe,eq}/\Omega_{ce,eq}$.

Turning next to EMIC wave generation, the dispersion relation of field-aligned EMIC waves is \cite{bookStix62}
\begin{equation}
\frac{{k^2 c^2 }}{{\omega ^2 }} \approx  1- \frac{{\Omega _{pe}^2 }}{{\omega \Omega _{ce} }} - \frac{{\Omega _{pe}^2 }}{\omega }\frac{{m_e }}{{m_p }}\left( {\frac{{\eta _H }}{{\omega  - \Omega _{cp} }} + \frac{{\eta _{He} }}{{\omega  - \Omega _{cp} /4}} + \frac{{\eta _O }}{{\omega  - \Omega _{cp} /16}}} \right)
\label{eq04}    
\end{equation}
where $\eta_{H}$, $\eta_{He}$, $\eta_{O}$ are the relative concentrations of protons, helium ions, and oxygen ions, respectively (with $\eta_H+\eta_{He}+\eta_O=1$), and $\Omega_{cp}=\Omega_{ce}m_e/m_p$ is the proton gyrofrequency ($m_e$ and $m_p$ are the electron and proton mass, respectively). For a purely proton-electron plasma, equation (\ref{eq04}) can be rewritten as:
\begin{equation}
\omega  = \Omega _{cp} \left( \lambda  \right) \cdot \left( {\frac{{k\left( \lambda  \right)c}}{{\Omega _{pp} \left( \lambda  \right)}}} \right)^2  \cdot \left( { - \frac{1}{2} + \sqrt {\frac{1}{4} + \left( {\frac{{\Omega _{pp} \left( \lambda  \right)}}{{k\left( \lambda  \right)c}}} \right)^2 } } \right)
\label{eq05}    
\end{equation}
or
\[
k\left( \lambda  \right) = \frac{\omega }{c}\sqrt {1 + \frac{{\Omega _{pp}^2 \left( \lambda  \right)}}{{\Omega _{cp} \left( \lambda  \right) \cdot \left( {\Omega _{cp} \left( \lambda  \right) - \omega } \right)}}} 
 \approx \frac{\omega }{c}\frac{{\Omega _{pe} \left( \lambda  \right)}}{{\Omega _{ce} \left( \lambda  \right)}}\sqrt {\frac{{m_p }}{{m_e }}} \left( {1 - \frac{\omega }{{\Omega _{cp} \left( \lambda  \right)}}} \right)^{ - 1/2} 
\]
where $\Omega_{pp}^2=\Omega_{pe}^2m_e/m_p$. The resonance condition of equation (\ref{eq02}) can be rewritten for EMIC waves as
\begin{equation}
\gamma\omega  - k\left( \lambda  \right)p_\parallel\left( \lambda  \right)/m_e  = -\Omega _{ce} \left( \lambda  \right) \label{eq06}    
\end{equation}

\begin{figure*}
\centering
\includegraphics[width=0.95\textwidth]{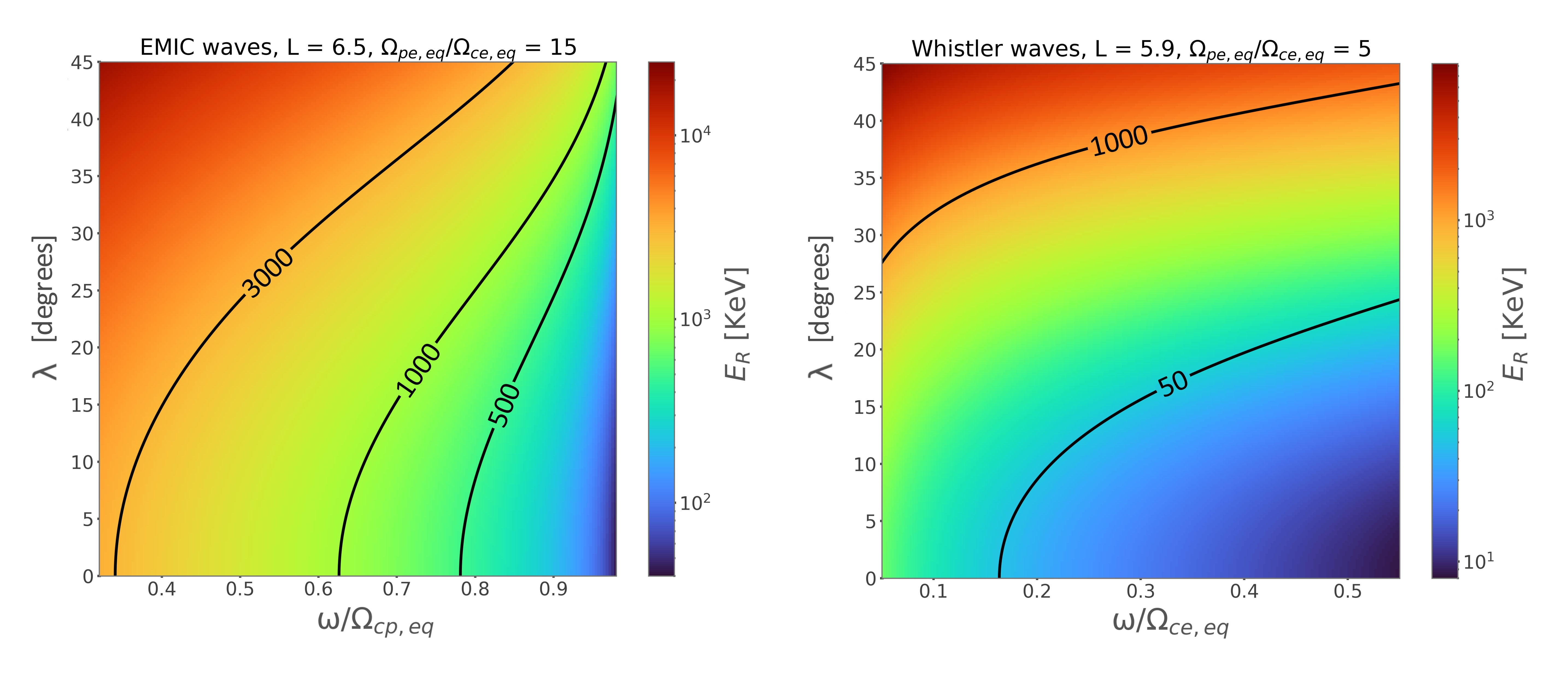}
\caption{Resonance energy as a function of magnetic latitude $\lambda$ and wave frequency $\omega$ for EMIC waves (left panel) and whistler-mode waves (right panel). The parameters for calculation are shown at the top of each panel; black lines represent contours of constant energies that approximately coincide with upper and lower bounds of ELFIN precipitation observations.
\label{fig6}}
\end{figure*}

Using these relations, we obtain resonance energies as a function of magnetic latitude and wave frequency, as displayed in Figure \ref{fig6}. For EMIC wave calculations we set $\Omega_{pe,eq}/\Omega_{ce,eq}=15$ \cite<typical for EMIC wave generation region outside the plasmasphere, see>{Zhang16:grl}, while for those of whistler-mode waves we set $\Omega_{pe,eq}/\Omega_{ce,eq}=5$ \cite<typical for whistler-mode wave generation region within the dawn flank, see>{Glauert&Horne05,Agapitov19:fpe}. Calculations for EMIC waves (left panel) show that to provide precipitation of both sub-relativistic ($<500$ keV) and accompanying relativistic ($\geq1$ MeV) electrons, as observed by ELFIN in S\#1 (see Fig. \ref{fig3}), the waves would likely need to be at a very high-frequency, with $\omega/\Omega_{cp,eq}>0.8$ \cite<see also>{Ukhorskiy10,Denton19,Bashir22:grl}. Such a high-frequency portion of EMIC wave spectra is indeed observed around equator \cite<see, e.g.,>{Zhang16:grl} and may provide the necessary precipitating-to-trapped flux ratio to induce the sub-relativistic precipitation associated with EMIC-driven relativistic electron precipitation \cite<see>{Angelopoulos23:ssr,Capannolo19:EMICSpatialExtent}. Two additional factors may facilitate such sub-relativistic precipitation by EMIC waves: (1) enhanced plasma density with $\Omega_{pe,eq}/\Omega_{ce,eq}$ exceeding the nominal (model) values \cite{Summers&Thorne03,Summers07:rates} and/or (2) non-resonant electron scattering occurring below the minimum resonance energy \cite<effective for short EMIC wavepackets, see>{Chen16:nonresonant,An22:prl,Grach&Demekhov23:theory}. Both these factors may contribute to the precipitating electron spectra that demonstrate a weak (precipitating-to-trapped ratio of $\sim1/10$) but finite precipitation down to $50$ keV, as seen in our first event.

Concerning whistler-mode waves, the precipitation of relativistic electrons suggests a large local $\Omega_{ce}/\Omega_{pe}$ \cite{Summers07:theory}, indicating that such waves should propagate up to high latitudes. This is indeed the case for the second event, as the precipitating-to-trapped electron flux ratio is approximately 1 \cite<i.e., at the strong diffusion limit, see>{Kennel69} for energies up to $0.9$ MeV (Figure \ref{fig6}, right panel shows that resonant latitudes are $\sim 40^\circ$ for such energies and typical wave frequency $\omega/\Omega_{ce,eq}\sim 0.3$ \cite{Li11,Agapitov18:jgr}). Empirical wave intensity models, such as those of \citeA{Agapitov18:jgr} and \citeA{Wang&Shprits19:latitudes}, predict that wave intensity should decrease with increasing magnetic latitude (i.e. farther away from the equator). This possible wave damping \cite<likely due to Landau resonance with suprathermal electrons; see>{Bell02,Bortnik07:landau,Chen13} prevents effective scattering of relativistic electrons. Thus, two possible scenarios can explain the observed electron precipitation of the second event. The first scenario assumes that electrons are scattered by whistler-mode waves ducted within a small-scale density perturbation \cite{Hosseini21:ducts,Ke21:ducts,Chen21:ducting,Shen21:grl:ducts} that can trap waves and prevent their damping \cite<see>[for discussion of the wave ducting effect on electron scattering energies]{Artemyev21:jgr:ducts,Chen22:microbursts}. The second scenario assumes that the electrons are scattered by very oblique whistler-mode waves resonating with electrons near the equator via high-order resonance: $\gamma\omega  - k\left( \lambda  \right)p_\parallel\left( \lambda  \right)/m_e  = n\Omega _{ce} \left( \lambda  \right)$  with $|n|>1$ \cite<e.g.,>{Lorentzen01,Mourenas12:JGR,Artemyev16:ssr}. Such waves can precipitate relativistic electrons even at low latitudes \cite<see examples in>{Gan23:grl_elfin}, but require additional populations of field-aligned suprathermal electron streams to supress Landau damping and thus allow very oblique wave generation \cite<e.g.,>{Mourenas15,Li16,Artemyev&Mourenas20:jgr,Sauer20}. Therefore, to explain the dawn-flank relativistic electron precipitation as driven by interplanetary shock wave impact, we would need to incorporate either strong equatorial density gradients formed by convection electric fields, which could duct whistler-mode waves, or an ionospheric outflow of secondary (suprathermal, $\sim 100$ eV) electrons \cite{Khazanov14:outflow,Khazanov22:HeatFluxWhistler} to provide the conditions for very oblique whistler-mode wave generation.

\section{Summary}\label{sec:conclusions}

In this study we explore the chain of events leading to relativistic electron precipitation during two different cases when comprehensive observations were available from the solar wind, the magnetosphere, and the ionosphere. The two events are each characterized by strong solar wind drivers that impact and compress the magnetosphere, triggering intense geomagnetic activity and electromagnetic wave intensification and ultimately culminating in distinct forms of electron precipitation. Through the combined multi-point observations of ARTEMIS, THEMIS,  and the geosynchronous, low-altitude ELFIN satellites, we synthesize an explanatory model of the sequence of events that lead to the observed characteristics of precipitation. Although the effects of EMIC and whistler-mode wave activity enhancements due to solar wind transient impacts have been well explored in various previous studies \cite{Yue17:whistler&solarwind, Blum21:emic&shock, Jin22:chorus&shock, Xue22:emic&swpressure, Li22:emic&shock, Yan23:emic&IS, Zuxiang23:ws_emic}, the two events in this paper arguably represent the first direct observations of relativistic electron losses as induced by solar wind-driven waves. 

In the first event, a corotating interaction region, containing an interplanetary shock and multiple rotational discontinuities of the solar wind, compresses the magnetosphere and drives a prolonged moderate storm with multiple substorm injections. Such injections are known to be responsible for hot ion transport into the inner magnetosphere, where the injected ion population consequently drives EMIC wave generation. We have presented low-altitude observations of resultant EMIC wave-driven strong losses (precipitating-to-trapped flux ratio reaches one) of relativistic electrons, spanning energies of $300$ keV to $\sim2$ MeV within a wide latitudinal ($L$-shell) range. In the second event an interplanetary coronal mass ejection, with a strong preceding interplanetary shock, impacts the magnetosphere and significantly compresses it ($Sym-H$ reaches $40$ nT). Such compression is known to drive whistler-mode waves, and we have presented low-altitude observations of intense whistler wave-driven electron precipitation, encompassing a wide energy range (from 50 to 700 keV) but very localized span of latitudes ($L$-shells). We have examined resonance conditions and cold plasma dispersion relations in order to evaluate the expected characteristics of waves capable of producing each unique precipitation event. Using these calculations, we have discussed plausible physical factors and scenarios which could foster the proper conditions for the latitudinal distribution of these waves. 

This study was largely built around ELFIN's low-altitude measurements of electron precipitation, and further investigations would benefit from the incorporation of additional near-equatorial spacecraft observations which could directly identify specific wave modes and their drivers (e.g., anisotropic ion and electron populations). Moreover, a combination of global magnetohydrodynamic (MHD) and test-particle simulations, outside the scope of this study, would be needed to verify solar wind structure impact as the main trigger for electron precipitation \cite<see discussion in>{Ukhorskiy22:NatSR, Chan23:frontiers}. The application of such models is left for future work.

\acknowledgments
We acknowledge support from NASA contract NAS5-02099. Work at Los Alamos National Laboratory was performed under the auspices of the United States Department of Energy. Q.M. would like to acknowledge the NASA grant 80NSSC20K0196 and NSF grant AGS-2225445.  A.V.A and X.-J.Z. acknowledge support from the NASA grants 80NSSC23K0108, 80NSSC23K0403, 80NSSC21K0729, and from the NSF grant 2021749. V.A. and A.R. also acknowledge support from NSF grants AGS-1242918, AGS-2019950.

We are grateful to NASA's CubeSat Launch Initiative for ELFIN's successful launch. We acknowledge early support of the ELFIN project by the AFOSR, under its University Nanosat Program; by the UNP-8 project, contract FA9453-12-D-0285; and by the California Space Grant program. We acknowledge the critical contributions of the numerous volunteer ELFIN team student members.  We acknowledge the GEO-KOMPSAT-2A magnetometer team (Ulrich Auster, Dragos Constantinescu; Institut für Geophysik und Extraterrestrische Physik, Technische Universität Braunschweig, Braunschweig, Germany) for their high quality dataset. We also acknowledge the support of NASA contract NAS5-02099 for the use of data from the THEMIS Mission, specifically K. H.Glassmeier, U. Auster, and W. Baumjohann for the use of FGM data (provided under the lead of the Technical University of Braunschweig and with financial support through the German Ministry for Economy and Technology and the German Center for Aviation and Space (DLR) under contract 50 OC 0302) and C. W. Carlson, J. P. McFadden for the use of ESA data.

\subsection*{Open Research}
\noindent ELFIN data is available at https://data.elfin.ucla.edu/\\
\noindent THEMIS\&ARTEMIS data is available at http://themis.ssl.berkeley.edu.\\
\noindent Sym-H and AE indexes were downloaded from https://supermag.jhuapl.edu/\\
\noindent GEO-KOMPSAT-2A (SOSMAG) data is made available via ESA's Space Safety Programme and its provision forms part of the ESA Space Weather Service System at https://swe.ssa.esa.int/\\ \noindent
\noindent Data access and processing was done using SPEDAS V3.1, see \citeA{Angelopoulos19}.

%% ---------------------------------------------------------------%%
%% References and Citations %%

%\bibliography{full,addon}

\end{document}